\documentclass[aps,prd,nofootinbib,superscriptaddress,10pt]{revtex4}
\usepackage{graphicx}
\usepackage{dcolumn}
\usepackage{amsmath}
\usepackage{epsfig}
\usepackage{colordvi}
\usepackage{color}
\usepackage{hhline}

\begin{document}
\title{\large \bf \boldmath
Measurement of the  $e^+e^- \to \pi^0\gamma$ cross section
in the energy range 1.075--2 GeV at SND}

\author{M.~N.~Achasov}
\author{A.~Yu.~Barnyakov}
\author{K.~I.~Beloborodov}
\author{A.~V.~Berdyugin}
\author{D.~E.~Berkaev}
\affiliation{Budker Institute of Nuclear Physics, SB RAS, Novosibirsk, 630090,
Russia}
\affiliation{Novosibirsk State University, Novosibirsk, 630090, Russia}
\author{A.~G.~Bogdanchikov}
\author{A.~A.~Botov}
\affiliation{Budker Institute of Nuclear Physics, SB RAS, Novosibirsk, 630090,
Russia}
\author{T.~V.~Dimova}
\author{V.~P.~Druzhinin}
\author{V.~B.~Golubev}
\author{L.~V.~Kardapoltsev}
\affiliation{Budker Institute of Nuclear Physics, SB RAS, Novosibirsk, 630090,
Russia}
\affiliation{Novosibirsk State University, Novosibirsk, 630090, Russia}
\author{A.~S.~Kasaev}
\affiliation{Budker Institute of Nuclear Physics, SB RAS, Novosibirsk, 630090,
Russia}
\author{A.~G.~Kharlamov}
\affiliation{Budker Institute of Nuclear Physics, SB RAS, Novosibirsk, 630090,
Russia}
\affiliation{Novosibirsk State University, Novosibirsk, 630090, Russia}
\author{I.~A.~Koop}
\affiliation{Budker Institute of Nuclear Physics, SB RAS, Novosibirsk, 630090,
Russia}
\affiliation{Novosibirsk State University, Novosibirsk, 630090, Russia}
\affiliation{Novosibirsk State Technical University, Novosibirsk, 
630092, Russia}
\author{A.~A.~Korol}
\affiliation{Budker Institute of Nuclear Physics, SB RAS, Novosibirsk, 630090,
Russia}
\affiliation{Novosibirsk State University, Novosibirsk, 630090, Russia}
\author{D.~P.~Kovrizhin}
\author{S.~V.~Koshuba}
\affiliation{Budker Institute of Nuclear Physics, SB RAS, Novosibirsk, 630090,
Russia}
\author{A.~S.~Kupich}
\affiliation{Budker Institute of Nuclear Physics, SB RAS, Novosibirsk, 630090,
Russia}
\affiliation{Novosibirsk State University, Novosibirsk, 630090, Russia}
\author{R.~A.~Litvinov}
\author{K.~A.~Martin}
\affiliation{Budker Institute of Nuclear Physics, SB RAS, Novosibirsk, 630090,
Russia}
\author{N.~A.~Melnikova}
\author{N.~Yu.~Muchnoi}
\affiliation{Budker Institute of Nuclear Physics, SB RAS, Novosibirsk, 630090,
Russia}
\affiliation{Novosibirsk State University, Novosibirsk, 630090, Russia}
\author{A.~E.~Obrazovsky}
\author{E.~V.~Pakhtusova}
\affiliation{Budker Institute of Nuclear Physics, SB RAS, Novosibirsk, 630090,
Russia}
\author{K.~V.~Pugachev}
\author{Yu.~A.~Rogovsky}
\author{A.~I.~Senchenko}
\author{S.~I.~Serednyakov}
\author{Z.~K.~Silagadze}
\author{P.~Yu.~Shatunov}
\author{Yu.~M.~Shatunov}
\affiliation{Budker Institute of Nuclear Physics, SB RAS, Novosibirsk, 630090,
Russia}
\affiliation{Novosibirsk State University, Novosibirsk, 630090, Russia}
\author{D.~A.~Shtol}
\affiliation{Budker Institute of Nuclear Physics, SB RAS, Novosibirsk, 630090,
Russia}
\author{D.~B.~Shwartz}
\affiliation{Budker Institute of Nuclear Physics, SB RAS, Novosibirsk, 630090,
Russia}
\affiliation{Novosibirsk State University, Novosibirsk, 630090, Russia}
\author{I.~K.~Surin}
\author{ Yu.~V.~Usov}
\affiliation{Budker Institute of Nuclear Physics, SB RAS, Novosibirsk, 630090,
Russia}
\author{A.~V.~Vasiljev}
\author{V.~V.~Zhulanov}
\affiliation{Budker Institute of Nuclear Physics, SB RAS, Novosibirsk, 630090,
Russia}
\affiliation{Novosibirsk State University, Novosibirsk, 630090, Russia}

\begin{abstract}
The process $e^+e^- \to \pi^0\gamma$ is studied  with the SND detector 
at the VEPP-2000 $e^+e^-$ collider. Basing on data with an integrated 
luminosity of 41 pb$^{-1}$ recorded in 2010-2012 we measure
the $e^+e^- \to \pi^0\gamma$ cross section in the center-of-mass energy
range from 1.075 up to 2 GeV. In the range 1.4-2.0 GeV the process
$e^+e^- \to \pi^0\gamma$ is studied for the first time.
\end{abstract}

\maketitle

\section{Introduction}
In this paper we begin to study the $e^+ e^- \to\pi^0 \gamma$ process
in experiments at the VEPP-2000 $e^+e^-$-collider~\cite{vepp2000}.
The previous most accurate measurements of this process
were performed in experiments at the VEPP-2M $e^+e^-$ collider
with the SND~\cite{snd1,snd2,snd3} and CMD-2~\cite{cmd} detectors.

The experimental data on the process $e^+ e^- \to \gamma^\ast\to\pi^0 \gamma$
is needed, in particular, for development of phenomenological models
of the $\pi^0\gamma^{(\ast)}\gamma^{(\ast)}$ transition form factor 
used in the calculation of the hadronic
light-by-light contribution to the value of the muon anomalous magnetic
moment (see, for example, \cite{lbl}).

The energy dependence of the $e^{+}e^{-}\rightarrow \pi ^{0}\gamma$ cross 
section is well described by the vector meson dominance (VMD) model.
From the fit to the cross-section data the widths of radiative 
decays of vector mesons $\rho(770)$, $\omega(782)$, and $\phi(1020)\to
\pi^0\gamma$ are extracted. These parameters are widely used in
phenomenological models, in particular, to fix the vector-meson quark 
content. It should be noted that currently the systematic 
uncertainties of some VMD-model parameters, such as the $\phi(1020)\to
\pi^0\gamma$ width, and the relative phases between the 
$\rho(770)$, $\omega(782)$, and $\phi(1020)$ resonance
amplitudes are determined by uncertainty in the contributions of
excited vector states. The evidence of the nonzero contribution
of $\rho(1450)$ and $\omega(1420)$ resonances was obtained
in Refs.~\cite{snd3,cmd}, where the $e^{+}e^{-}\rightarrow \pi ^{0}\gamma$
cross section was measured up to the center-of-mass energy $\sqrt{s}=1.4$ GeV. 

In this work, in the experiment with the SND detector at VEPP-2000, we
repeat the measurement of the $e^{+}e^{-}\rightarrow \pi ^{0}\gamma$
cross section in the energy range $\sqrt{s}=1.075-1.4$ GeV and explore 
a new region, between 1.4 and 2 GeV.

\section{Experiment} \label{Exper}
   SND~\cite{snd} is a general-purpose non-magnetic detector. 
Its main part is a spherical three-layer calorimeter containing 1640 
NaI(Tl) crystals. The calorimeter covers a solid angle of 95\% of $4\pi$; its 
thickness for particles coming from the collider interaction region is
$ 13.4\, X_{0} $. The calorimeter energy resolution for
photons is $\sigma_{E}/E_\gamma=4.2\%/\sqrt[4]{E_\gamma({\rm GeV)}}$, 
the angular resolution $\simeq 1.5^\circ$. Directions of charged particles are
measured by a nine-layer cylindrical drift chamber.
Outside the calorimeter a muon detector is located, which consists of
plastic scintillation counters and drift tubes. In this analysis
the muon detector is used as a cosmic-ray veto.

The analysis presented in this paper is based on data with an integrated 
luminosity of 41 pb$^{-1}$ recorded with the SND detector in 2010--2012 in 
the c.m. energy range $\sqrt{s}=1.075-2.000$ GeV. The data were collected 
in 56 energy points during several c.m. energy scans.
Due to relatively small statistics of $e^+e^-\to \pi^0 \gamma$ events 
in the energy region under study the energy points are merged
into 11 energy intervals listed in Table~\ref{allres}.

For simulation of signal events, we use an event generator, 
which takes into account radiative corrections to the initial particles
calculated according to Ref.~\cite{FadinRad}. The angular distribution of 
additional photons radiated by the initial electron and positron is simulated 
according to Ref.~\cite{BoneMartine}.
The event generator for the process $e^+e^-\to \gamma\gamma(\gamma)$ used 
for normalization is based on Ref.~\cite{Berends}. The theoretical 
uncertainty of the $e^+e^-\to \gamma\gamma$ cross section calculation
is estimated to be 1\%. 
Interactions of the generated particles with the detector materials are 
simulated using GEANT4 software~\cite{geant4}. The simulation takes into 
account variations of experimental conditions during data taking, in 
particular, dead detector channels, and beam-generated background. Due to the 
beam background, some part of data events contains spurious tracks in the
drift chamber and photons. 
To take this effect into account, simulation uses special background
events recorded during data taking with a random trigger, which are 
superimposed on simulated events.

\section{Event selection\label{evsel}}
In this analysis, we simultaneously select three-photon events of
the process under study 
$e^+e^-\to\pi^0\gamma\to 3\gamma$
and two-photon events of the process
$e^+e^-\to \gamma\gamma$ used for normalization.
Some selection criteria, such as absence of
charged tracks and extra photon in an event, and the muon-system veto, 
are common for both processes. So, 
systematic uncertainties associated with these 
criteria cancel as a result of the normalization.

The preliminary selection criteria for two- and three-photon 
events are following:
\begin{itemize}
\item No charged tracks are reconstructed in the drift chamber. The number of
hits in the drift chamber is less than four.
\item The total energy deposition in the calorimeter is larger than
$0.65\sqrt{s}$.
\item The total event momentum calculated using energy
deposition in the calorimeter crystals is less than $0.3\sqrt{s}$.
\item No signal in the muon system.
\end{itemize}

Two-photon events of the process $e^+e^-\to \gamma\gamma$ are selected with the
following additional criteria. There are exactly two photons in
an event. Their energies is required to be larger than 
with $0.3\sqrt{s}$. The azimuthal and polar angles of 
the photons satisfy the conditions 
$| |\phi_1-\phi_2| - 180^{\circ}| < 15^{\circ}$,
$|\theta_1+\theta_2 - 180^{\circ}| < 25^{\circ}$,
and
$(180^{\circ}-|\theta_1-\theta_2|)/2 > 45^{\circ}  $.
The integrated luminosity measured for each energy interval is listed in 
Table~\ref{allres}. The systematic uncertainty on the luminosity measurement is 
estimated to be 1.4\% and includes the theoretical error of cross section 
calculation (1\%) and uncertainty associated with a difference between data 
and simulation in photon angular and energy resolutions (1\%). The latter
uncertainty was carefully studied in Refs.~\cite{snd3,sndompi}.

The $e^+e^-\to \pi^0\gamma$ candidate event must have exactly three 
reconstructed photons. The energy of the less-energetic photon in an event 
$E_{\gamma,min}$ is required to be greater than 100 MeV. For these events we 
perform a kinematic fit with four constraints of energy and momentum balance.
The condition on the $\chi^2$ of the kinematic fit $\chi^2_{3\gamma}<20$ 
is applied.
For further selection we use the fitted parameters of the three photons.
Their polar angles are required to be in the range
$36^{\circ }<\theta_\gamma<144^{\circ }$, and their energy be
larger than $0.075\sqrt{s}$. The main source of background for the process
under study is $e^+e^-\to 3\gamma$ events. They are strongly suppressed
by the conditions on $E_{\gamma,min}$ and $\theta_\gamma$.
We calculate the mass recoiling against the most energetic
photon in an event $M_{\rm rec}$ and select events with 
$80<M_{\rm rec}<190$ MeV/$c^2$.

\begin{figure}
\includegraphics[width=0.4\textwidth]{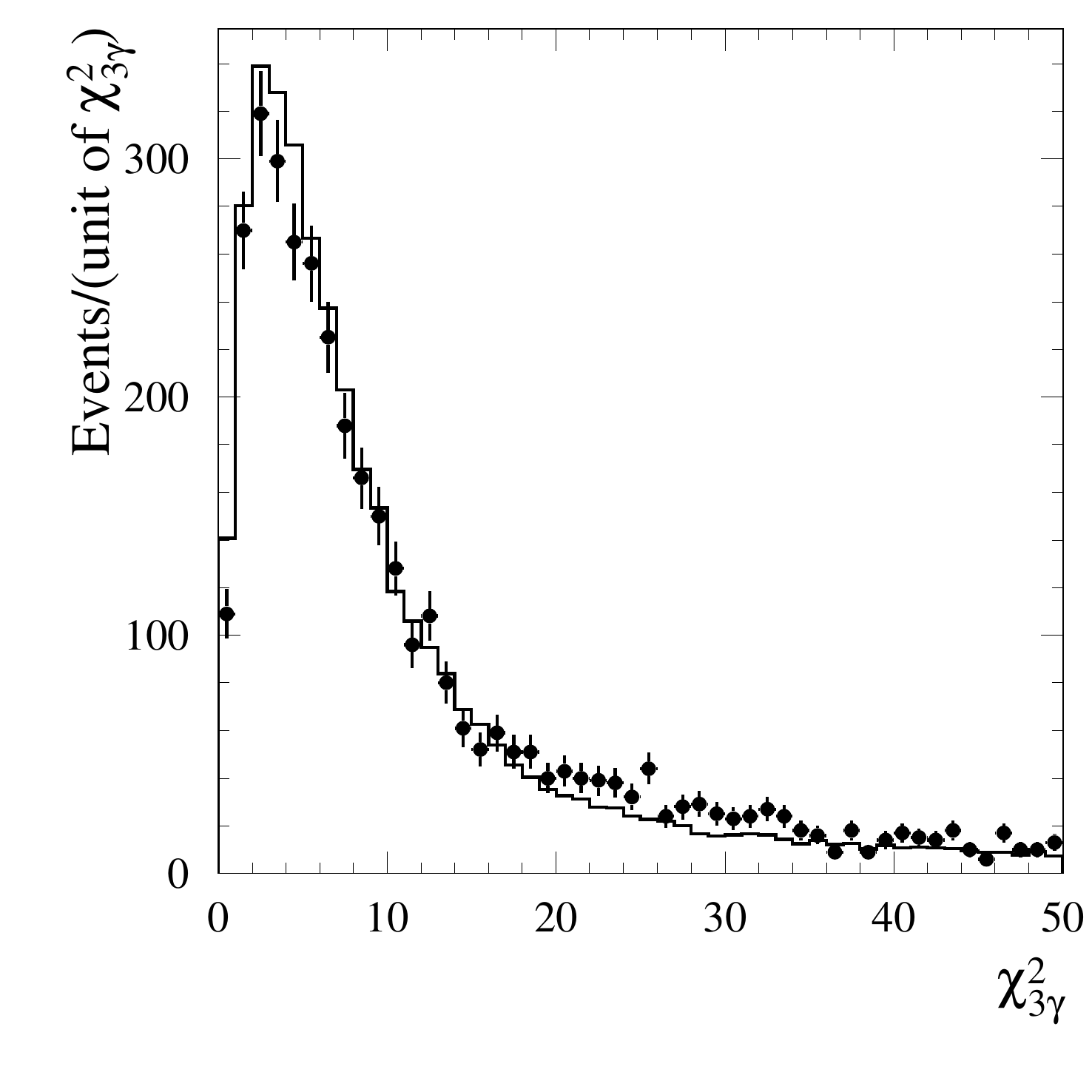}
\caption{The $\chi^2_{3\gamma}$ distribution for data (points with
error bars) and simulated $e^+e^-\to 2\gamma (\gamma)$ events (histogram).
The simulated distribution is normalized to the number of expected events.
\label{fig1}}
\end{figure}

The distribution of $\chi^2$ of the kinematic fit ($\chi^2_{3\gamma}$) for
data $e^+e^-\to \pi^0\gamma$ candidate events is shown in Fig.~\ref{fig1}
as points with error bars. 
The histogram in Fig.~\ref{fig1} represents the expected distribution for 
$e^+e^-\to 2\gamma (\gamma)$ events obtained using MC simulation.
The number of events in the histogram is calculated as
$\sum_i 
N_{2\gamma,i}^{\rm data}(N_{3\gamma,i}^{\rm MC}/N_{2\gamma,i}^{\rm MC})$,
where $N_{2\gamma,i}^{\rm data}$ and $N_{2\gamma,i}^{\rm MC}$ are the
numbers of selected $e^+e^-\to \gamma\gamma$ events in data and 
in $e^+e^-\to 2\gamma (\gamma)$ simulation, respectively, and 
$N_{3\gamma,i}^{\rm MC}$ is the number of three-photon events 
in the $e^+e^-\to 2\gamma (\gamma)$ simulation. It is seen that
most of the selected data events originate from the background 
process $e^+e^-\to 3\gamma$. 
The difference between the tails of the data and MC distributions
is due to the absence radiative corrections (extra photon radiation) to the 
$e^+e^-\to 3\gamma$ process in the $e^+e^-\to 2\gamma (\gamma)$ MC generator
used.

\section{Fitting the $M_{\rm rec}$ spectra\label{mrecfit}}
The number of signal events ($N_{\rm sig}$) is determined from
the fit to the $M_{\rm rec}$ spectrum by a sum of signal and background 
distributions. The signal distribution is described by a histogram obtained
from $e^+e^-\to \pi^0\gamma$ simulation. The simulation includes
radiation of an addition photon by the initial electron or positron. In
particular, the initial-state radiation (ISR) process 
$e^+e^-\to \omega\gamma \to \pi^0 \gamma\gamma$ is simulated. To correctly 
reproduce the rate of ISR events the event generator uses data on the the Born 
$e^+e^-\to \pi^0\gamma$ cross section obtained in this work and 
Ref.~\cite{snd3} and fitted in the frame of the 
VMD model. The $M_{\rm rec}$ spectrum for
simulated $e^+e^-\to \pi^0\gamma(\gamma)$ events is divided into
two parts: with $2E_{\rm ISR}/\sqrt{s}<0.4$ (signal) and  
$2E_{\rm ISR}/\sqrt{s}>0.4$ (ISR background), where $E_{\rm ISR}$ is
the energy of the extra photon emitted from the initial state.

\begin{figure}
\includegraphics[width=0.4\textwidth]{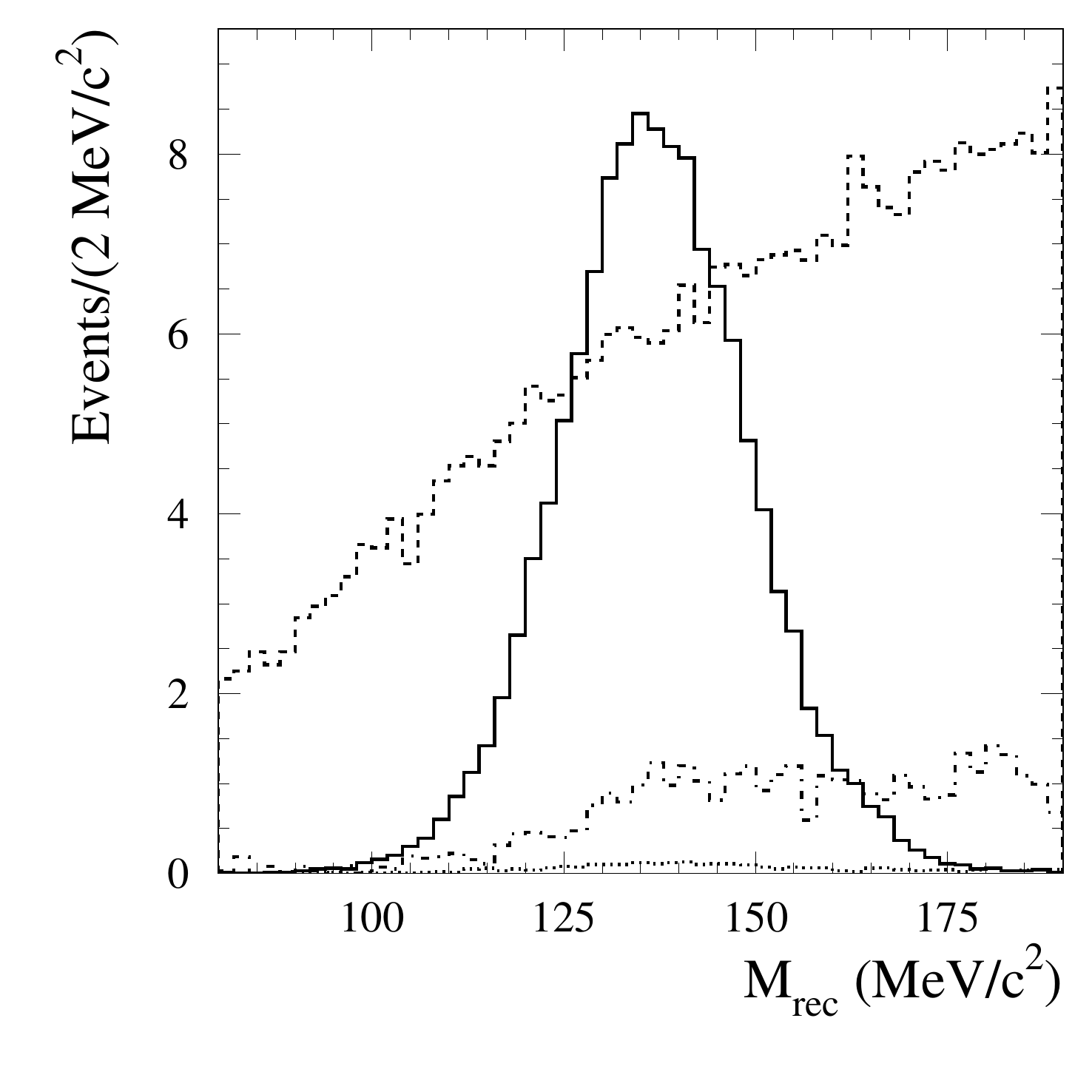}
\caption{The expected $M_{\rm rec}$ distributions (sum over all energy
points) for signal (solid histogram) and ISR (dotted histogram),
$e^+e^-\to 3\gamma$ (dashed histogram), and $e^+e^-\to \omega\pi^0$ 
(dash-dotted) backgrounds. The distribution for $e^+e^-\to 3\gamma$
events is multiplied by a factor of 0.1.
\label{fig2}}
\end{figure}

The other sources of background for the process under study are 
$e^+e^-\to 3\gamma$ and $e^+e^-\to \omega\pi^0\to \pi^0\pi^0\gamma$ events.
The latter background is estimated using MC simulation and 
cross-section data obtained in Ref.~\cite{sndompi}. Its fraction
does not exceed 2\% of the total background. The ISR background
fraction is about $10^{-3}$.
The shapes of the signal and background distributions obtained using
MC simulation are presented in Fig.~\ref{fig2}.
The simulation overestimates the number of background events 
by about 15\%. In the fit to the $M_{\rm rec}$ spectrum, the background 
distribution is a sum of the ISR, $3\gamma$, and $\omega\pi^0$
distributions plus a linear function.
The latter is needed to take into account a difference between 
data and simulation in the shape of the background distribution and in 
the number of background events.

\begin{figure*}
\includegraphics[width=0.4\textwidth]{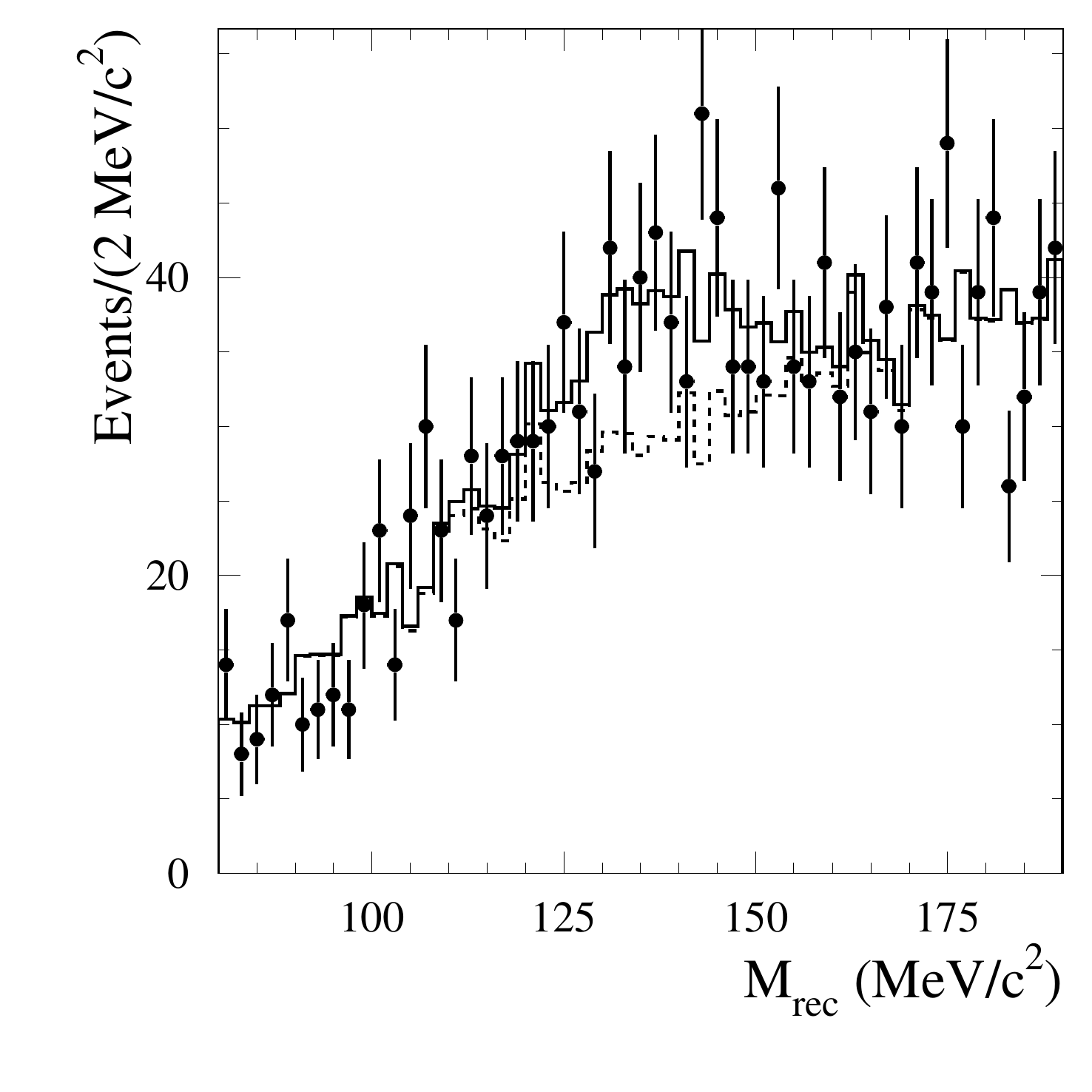}
\includegraphics[width=0.4\textwidth]{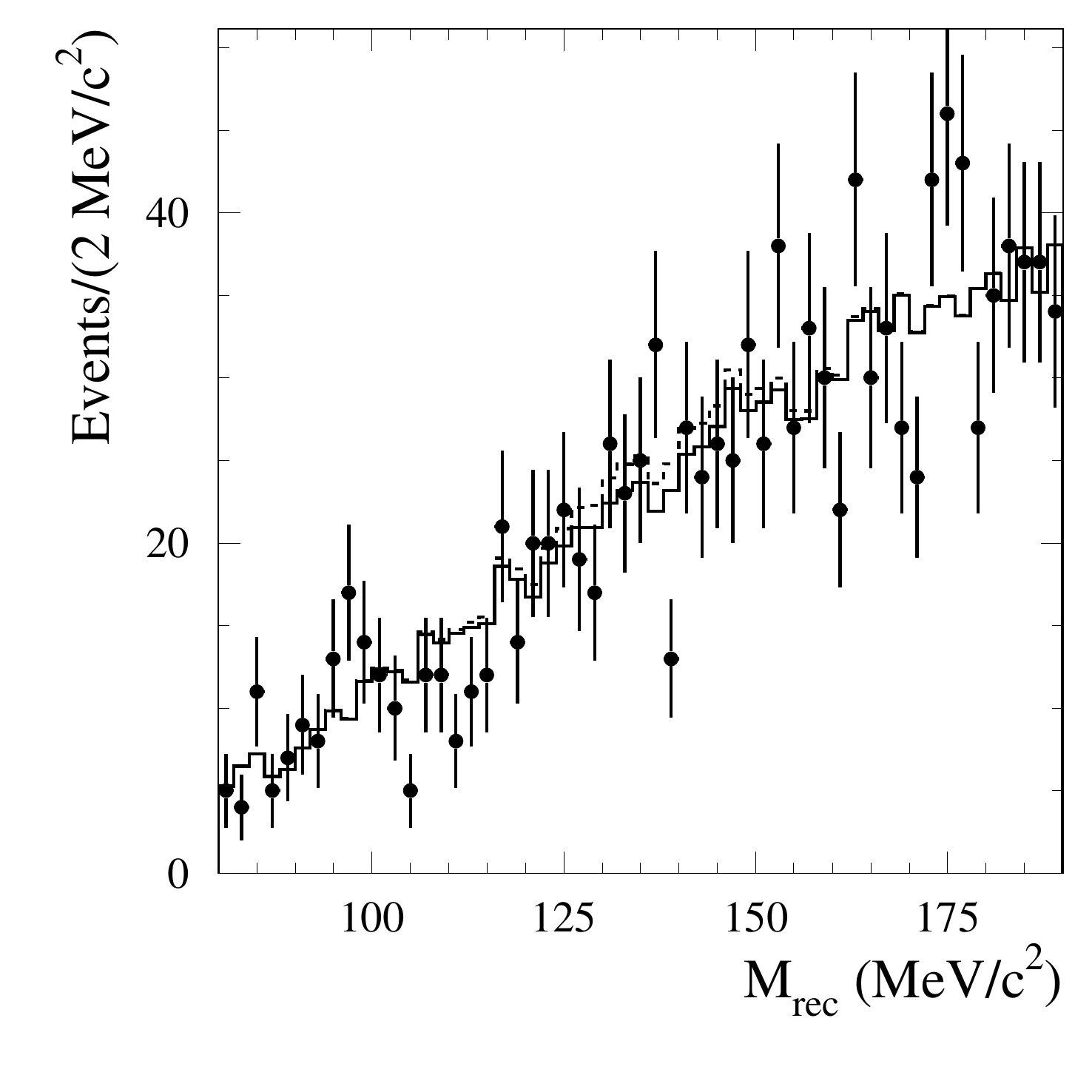}
\caption{The $M_{\rm rec}$ distribution for data events (points with error 
bars) with $\sqrt{s}=1075-1375$ GeV (left), and $\sqrt{s}=1400-2000$ MeV 
(right). The solid histogram represents the result of
the fit described in the text. The dashed histogram shows the fitted
background distribution.
\label{fig3}}
\end{figure*}

  The results of the fit in the energy regions $\sqrt{s}=1075-1375$ MeV
and  $\sqrt{s}=1400-2000$ MeV are shown in Fig.~\ref{fig3}.
The fitted number of signal events in these two regions are
$148 \pm 34$ and $-26 \pm 30$. We do not observe a $e^+e^-\to \pi^0\gamma$
signal at $\sqrt{s}=1400-2000$ MeV.
The obtained numbers of signal events for different energy intervals
are listed in Table~\ref{allres}.

\section{Detection efficiency and radiative corrections}
The visible cross section for the process $e^+e^- \to \pi^0\gamma$
is written as
\begin{equation}
\sigma_{vis}(s) = \int \limits_{0}^{x_{max}}\varepsilon_r(s, x)F(x,s)
\sigma(s(1-x))dx,
\label{viscrs}
\end{equation}
where $\sigma(s)$ is the Born cross section extracted
from the experiment, $F(x,E)$ is a so-called radiator function describing the
probability to emit from the initial state extra photons with the total energy
$x\sqrt{s}/2$~\cite{FadinRad}, $x_{max}=0.4$ (see Sec.~\ref{mrecfit}), and 
$\varepsilon_r(s, x)$ is the detection efficiency. The detection
efficiency is determined using MC simulation, as a function of $\sqrt{s}$ 
and $x=2E_{\rm ISR}/\sqrt{s}$. It is parametrized as 
$\varepsilon_r(s, x)=\varepsilon(s)g(s,x)$, where 
$\varepsilon(s)\equiv \varepsilon_r(s,0)$.
We use the approximation when all variations of experimental conditions 
(dead calorimeter channels, beam background, etc.) are accounted for in 
$\varepsilon(s)$, while $g(s,x)$ is a smooth function of $\sqrt{s}$. With this
parametrization, Eq.~(\ref{viscrs}) can be rewritten in the conventional form:
\begin{equation}
\label{viscrs2} \sigma_{vis} = \varepsilon(s)\sigma(s)(1+\delta(s)),
\end{equation}
where $\delta(s)$ is the radiative correction.

\begin{figure}
\includegraphics[width=0.4\textwidth]{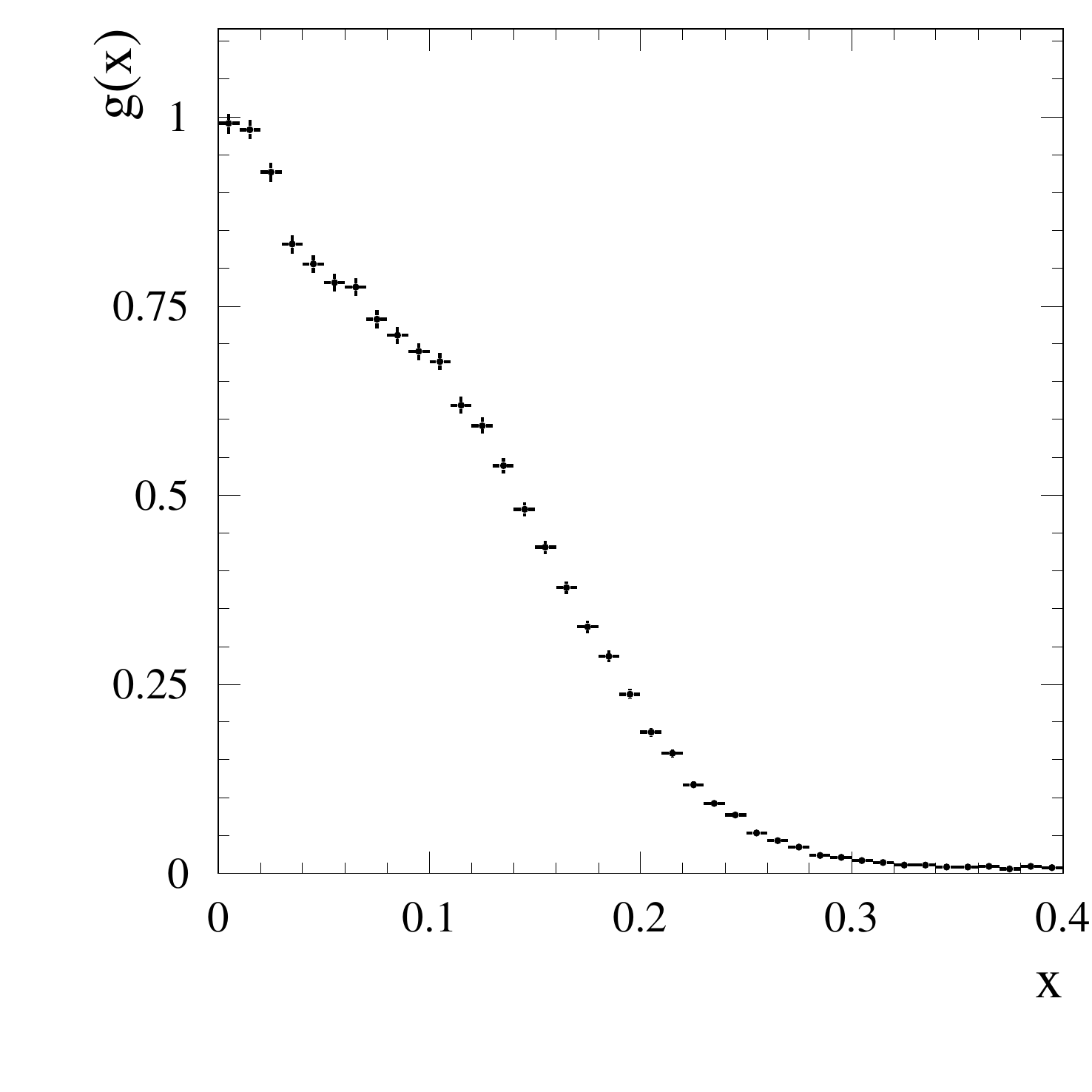}
\caption{The $x$ dependence of the detection efficiency obtained
from simulation for $\sqrt{s}=1675-2000$ MeV.
\label{fig4}}
\end{figure}

The functions $\varepsilon(s)$ and $g(s,x)$ are determined using MC
simulation. Dependence of the $g(s,x)$ shape on $s$ is not strong. 
The $x$ dependence of the detection efficiency is shown in Fig.~\ref{fig4}.
Two effects determine the shape of the function $g(x)$.
About 24\% of ISR photons are emitted into the angular range covered by
the SND calorimeter and may be detected and reconstructed if their energy 
higher than the photon reconstruction threshold 20 MeV. The sharp decrease 
of the efficiency at $x=0.02-0.04$ is due to requirement
that the number of reconstructed photons in an events is exactly 3, which
removes events with a reconstructed ISR photon. The further
decrease of the efficiency is due to the requirement of energy and
momentum balance in an event (condition $\chi^2_{3\gamma}<20$).

The values of $\varepsilon(s)$ for different energy intervals
are listed in Table~\ref{allres}. It decreases with increasing
$\sqrt{s}$ from 29\% at 1075 MeV to 14\% in the last energy interval
$\sqrt{s}=1870-2000$ MeV. The efficiency obtained using MC simulation is 
corrected (multiplied) by a factor of 1.013, which take into account the
difference between data and simulation in photon conversion in detector 
material before the tracking system (-0.3\%) and in the
shape of the $\chi^2_{3\gamma}$ distribution (1.6\%). 
The conversion probability is measured using $e^+e^-\to \gamma\gamma$ events.
To determine the efficiency correction for the $\chi^2_{3\gamma}$ cut and 
estimate systematic uncertainties due to imperfect 
simulation of the detector response for photons, we study simulated 
signal events and data events collected in 2013 in the maximum of the $\omega$ 
resonance, where the signal-to-background ratio is about 25. In particular, we
loosen the condition $\chi^2_{3\gamma}<20$ and vary the boundaries of the 
condition $36^{\circ }<\theta_\gamma<144^{\circ }$. The systematic uncertainty 
in the detection efficiency is estimated to be less than 2\%.

\begin{table}
\caption{The c.m.~energy ($E\equiv\sqrt{s}$), integrated luminosity ($L$), detection
efficiency ($\varepsilon$), number of selected signal events ($N_{\rm sig}$),
radiative-correction factor ($1+\delta$), measured Born cross section
($\sigma$). The radiative correction uncertainty is given in parentheses. 
For the cross section the first error is statistical, the
second is systematic.\label{allres}}
\begin{ruledtabular}
\begin{tabular}{cccccc}
 $E$, GeV & $L$, nb$^{-1}$ & $\varepsilon$, \% & $N_{\rm sig}$ & $1+\delta$ &
 $\sigma$, pb \\[0.3ex] \hline \\[-2.1ex]
 1075             &  569& 29.7& $ 21\pm 10$ & 1.32(6) & $ 93\pm45\pm4$ \\
 1119 (1100--1150)& 2043& 27.6& $ 28\pm 17$ & 1.02(6) & $ 48\pm29\pm3$ \\
 1200 (1175--1225)& 2398& 28.8& $ 38\pm 16$ & 0.92(2) & $ 60\pm26\pm2$ \\
 1284 (1250--1300)& 3154& 29.4& $ 43\pm 18$ & 0.92(2) & $ 50\pm21\pm2$ \\
 1353 (1325--1375)& 2666& 28.4& $ 16\pm 14$ & 0.94(3) & $ 23\pm19\pm1$ \\
 1425 (1400--1450)& 3290& 27.4& $-12\pm 14$ & 1.00(5) & $-13\pm15\pm1$ \\
 1509 (1475--1550)& 4762& 26.3& $-14\pm 15$ & 1.10(8) & $-11\pm13\pm1$ \\
 1607 (1575--1650)& 3593& 22.5& $ 11\pm 13$ & 1.06(16)& $ 12\pm16\pm2$ \\
 1705 (1675--1750)& 3895& 20.3& $ -1\pm 12$ & 0.88(14)& $ -1\pm13\pm2$ \\
 1804 (1760--1850)& 5432& 16.7& $  4\pm 14$ & 0.91(6) & $  4\pm15\pm1$ \\
 1926 (1870--2000)& 8955& 14.1& $ -9\pm 14$ & 0.92(3) & $ -8\pm13\pm1$ \\
\end{tabular}	   
\end{ruledtabular}
\end{table}

\section{Fit to cross section data}
To determine radiative corrections and calculate the Born cross section, the 
energy dependence of the measured visible cross section 
$\sigma_{vis,i}=N_{{\rm sig},i}/L_i$ is fitted with Eq.~(\ref{viscrs}).
The Born cross section is 
parametrized in the framework of the VMD model
as follows (see, for example, Ref.~\cite{FL})
\begin{eqnarray}
\sigma(s) & = & \frac{q_\gamma(s)^{3}}{s^{3/2}}
\left| \sum_{V} A_V(s) \right| ^{2},\label{vsum}\\
A_{V}(s) & = & \frac{m_{V}\Gamma _{V}e^{i\varphi_V}}
{m_V^2-s-i\sqrt{s}\Gamma _{V}(s)}
\sqrt{\frac{m_{V}^{3}}{q_\gamma(m_{V}^{2})^{3}}\sigma_V},\\
q_\gamma(s) & = & \frac{\sqrt{s}}{2}\left( 1-\frac{m^{2}_{\pi ^{0}}}{s}\right),
\end{eqnarray}
where $ m_{V} $ is the $ V $ resonance mass, $ \Gamma _{V}(s) $
is its energy-dependent width, $\Gamma _{V}\equiv\Gamma _{V}(m_{V}^2)$,
$\varphi_V$ is the interference phase, $\sigma_V$ is the cross section
at the resonance peak, which is related to the product of the 
branching fractions for the decays $V\to e^+e^-$ and $V\to\pi^0\gamma$: 
\begin{equation}
\sigma_V=\frac{12\pi}{m_V^2}B(V\to e^+e^-)B(V\to\pi^0\gamma).
\end{equation}
The sum in Eq.(\ref{vsum}) goes over the resonances of the $\rho$ and $\omega$
families, and $\phi(1020)$. To fix contributions of the $\rho(770)$, 
$\omega(782)$, and $\phi(1020)$ resonances, data obtained in this work are 
fitted simultaneously with $e^+e^-\to \pi^0\gamma$ data from  Ref.~\cite{snd3} 
obtained at $\sqrt{s}<1.4$ GeV. 

The detailed description of fit parameters is given in Ref.~\cite{snd3}. 
The contribution of the excited vector states $\omega(1420)$, $\rho(1450)$, 
$\omega(1650)$, and $\rho(1700)$ are parametrized by a sum of two effective 
resonances ($V^\prime$ and $V^{\prime\prime}$)
with masses of 1450 MeV and 1700 MeV. The $V^{\prime\prime}$ width is 
fixed at 315 MeV, the Particle Data Group (PDG) value for 
$\omega(1650)$~\cite{pdg}, while the $V^\prime$ width is a free fit parameter.
The fitted value of
$\sigma_{V^{\prime\prime}}=2\pm10$ pb is consistent with zero. The fitted $V^\prime$
parameters are $\sigma_{V^{\prime}}=16\pm10$ pb, and 
$\Gamma_{V^{\prime}}=480\pm180$ MeV. The latter value is in agreement
with the PDG values for $\omega(1420)$ and $\rho(1450)$~\cite{pdg}.

\begin{figure}
\includegraphics[width=0.4\textwidth]{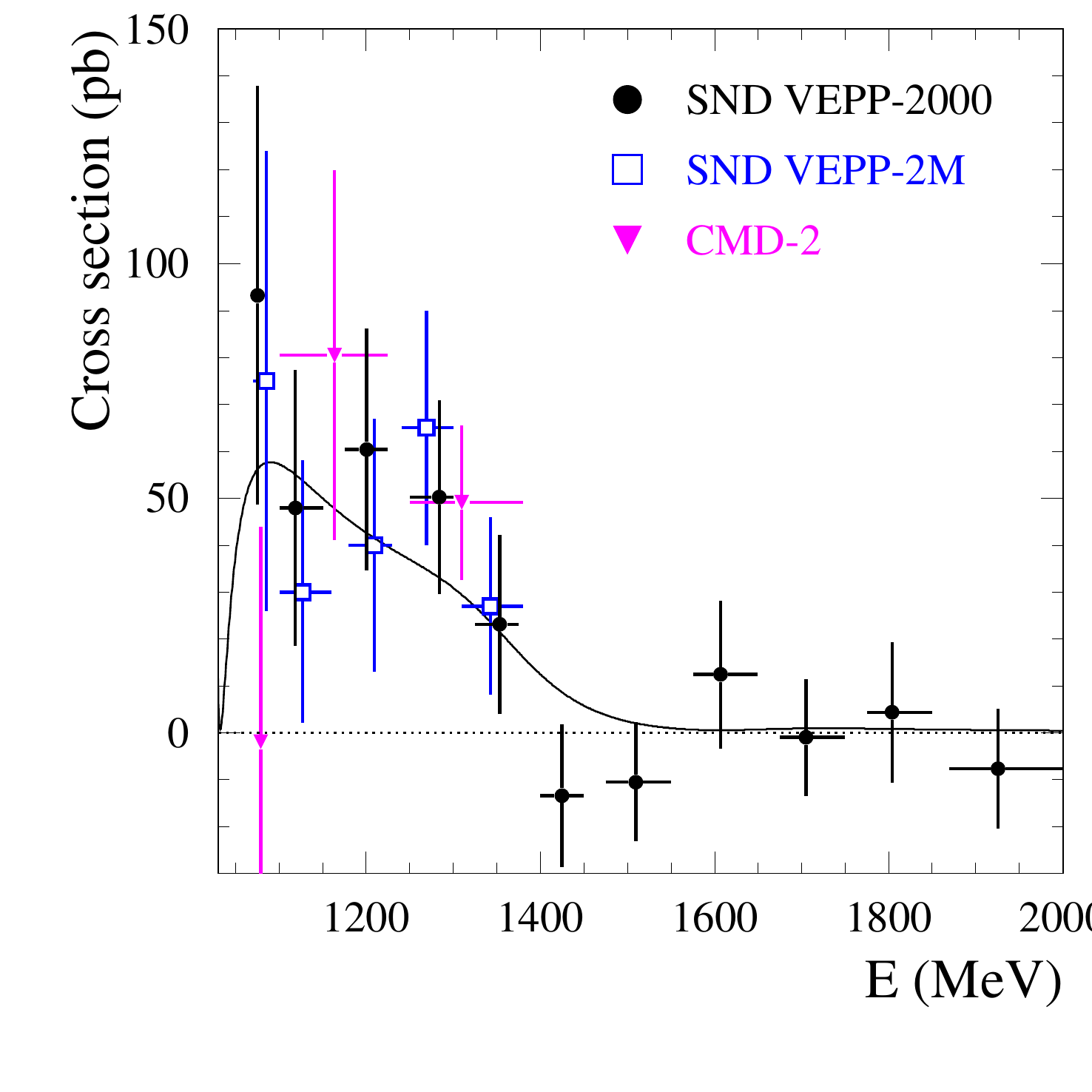}
\caption{The $e^+e^-\to \pi^0\gamma$ cross section measured in this work
in comparison with the previous measurements: SND~\cite{snd3},
and CMD-2 (2005)~\cite{cmd}. The curve is the result of the fit described 
in the text. \label{fig5}}
\end{figure}

As result of the fit we calculate the
radiative corrections and the experimental values of the Born cross section, 
which are listed in Table~\ref{allres}. The radiative corrections are also 
calculated with the fitted $V^{\prime}$ parameters varied within their 
uncertainties, and $m_{V^{\prime}}$ variation in the range $1450\pm50$ MeV.
The maximum deviation of the radiative correction from its nominal value is 
taken as an estimate of its uncertainty. It is shown in parentheses in the 
$(1+\delta)$ column of Table~\ref{allres}.
The quoted errors on the Born cross section are statistical and systematic.
The latter includes the uncertainties in luminosity (1.4\%), detection
efficiency (2\%), and radiative correction.
The Born cross section measured in this work is shown in Fig.~\ref{fig5} in 
comparison with the previous measurements~\cite{cmd,snd3}.

In the energy region below 1.4 GeV the obtained cross section is in 
agreement with the previous SND and CMD-2 measurements~\cite{snd3,cmd}.
The curve shown in Fig.~\ref{fig5} is the result of the fit to SND data
from the current analysis and Ref.~\cite{snd3}. To understand the significance
of the contribution of excited vector states, we perform a fit with 
$\sigma_{V^{\prime}}=\sigma_{V^{\prime\prime}}=0$. The difference of the 
$\chi^2$ values for this
fit and the nominal fit 10.9 corresponds to a significance of 3.3$\sigma$.
The $e^+e^-\to \pi^0\gamma$ cross section in the energy region $1.4-2.0$
GeV is consistent to zero. So, we do not observe the
$\omega(1650)$ and $\rho(1700)$ decays to $\pi^0\gamma$ at the
current level of statistics.

\section{Summary}
The cross section for the process
$e^+e^-\to\pi^0\gamma$  has been
measured in the energy range of 1.075--2 GeV with the SND detector 
at the VEPP-2000 $e^+e^-$ collider. Below 1.4 GeV the 
obtained cross section is about 50 pb in agreement with the previous 
SND and CMD-2 measurements~\cite{snd3,cmd}. To explain this cross section
value, the contribution of the $\omega(1420)$ and $\rho(1450)$ resonances
is required with a significance of $3.3\sigma$. In the region 1.4--2.0 GeV the
process $e^+e^- \to \pi^0\gamma$ has been studied for the first time. The
cross section in this region has been found to be consistent with zero within
the statistical errors of about 15 pb. 

Part of this work related to the photon reconstruction algorithm in the
electromagnetic calorimeter is supported
by the Russian Science Foundation (project No. 14-50-00080).

 \end{document}